\documentclass[a4paper, 12pt]{article}
\usepackage{amsfonts}

\usepackage{amsmath}

\input{tcilatex}

\begin{document}

\noindent {\LARGE The Nature of Information in Quantum }

\noindent {\LARGE Mechanics}

\begin{quote}
\textbf{Rocco Duvenhage}

Department of Mathematics and Applied Mathematics

University of Pretoria, 0002 Pretoria, South Africa.

e-mail: rocco@postino.up.ac.za

3 June 2002

\bigskip \noindent \textbf{Abstract: }A suitable unified statistical
formulation of quantum and classical mechanics in a $\ast $-algebraic
setting leads us to conclude that information itself is noncommutative in
quantum mechanics. Specifically we refer here to an observer's information
regarding a physical system. This is seen as the main difference from
classical mechanics, where an observer's information regarding a physical
system obeys classical probability theory. Quantum mechanics is then viewed
purely as a mathematical framework for the probabilistic description of
noncommutative information, with the projection postulate being a
noncommutative generalization of conditional probability. This view
clarifies many problems surrounding the interpretation of quantum mechanics,
particularly problems relating to the measuring process.
\end{quote}

\section*{1. INTRODUCTION}

There are several problems surrounding the interpretation of quantum
mechanics, mainly involving the measuring process: What does the collapse of
the wave function mean? What causes it? And so on. In this paper we argue
that many of these problems are essentially present in classical mechanics
as well. If one accepts that the nature of information in quantum mechanics
differs from that in classical mechanics in the way to be explained below,
then quantum mechanics does not introduce conceptual problems not already
present in the measuring process in classical mechanics.

In classical mechanics a measurement is nothing strange. It is merely an
event where the observer obtains information about some physical system. A
measurement therefore changes the observer's information regarding the
system. One can then ask: What does the change in the observer's information
mean? What causes it? And so on. These questions correspond to the questions
above, but now they seem tautological rather than mysterious, since our
intuitive idea of information tells us that the change in the observer's
information simply means that he has received new data, and that the change
is caused by the reception of the new data. We will see that the quantum
case is no different, except that the nature of information in quantum
mechanics differs from that in classical mechanics. We now first describe
the basic idea.

Let's say an observer has information regarding the pure state (the point in
phase space) of a classical system, but not necessarily complete information
(this is the typical case, since precise measurements are not possible in
practice). Now the observer performs a measurement on the system to receive
new data (for example he might have information regarding a particle's
position, now he measures the particle's momentum). The observer's
information after this measurement then differs from his information before
the measurement. In other words, a measurement ``disturbs'' the observer's
information.

In classical mechanics we know that an observer's information isn't merely
changed, but is actually increased by a measurement (assuming the
measurement provides data not already accounted for in the observer's
information). We will view this as an assumption regarding the nature of
information which does not hold in quantum mechanics. On an operational
level, this can be seen as the essential difference between quantum
mechanics and classical mechanics: In both quantum and classical mechanics
the observer's information is changed by a measurement if the measurement
provides new data, but in classical mechanics the observer's information
before the measurement is compatible with his information after the
measurement, while in quantum mechanics this is not necessarily the case.
(Also see Heisenberg,$^{(1)}$ p. 20.)

More precisely, we view the observer's information about a physical system
as a set of probabilities which he assigns to the outcomes of all
measurements which he can perform on the system. When the observer performs
a measurement to receive new data concerning the system, these probabilities
generally change. In the classical case the probabilities change according
to conditional probability, also known as the Bayes rule. Roughly this can
be seen as the minimum change necessary to bring the observer's set of
probabilities up to date with the result of the measurement. By the word
``compatible'' above, we mean that any probability which is zero just before
the measurement (say the probability for some particle to be in a certain
region in space), will still be zero just after the measurement. In this
sense the observer's information can only increase due to a measurement of a
classical system. In quantum mechanics however, we have the projection
postulate instead of the classical conditional probability. The projection
postulate can change a zero probability, and when this happens we can say
that the observer's information before the measurement has been \emph{%
invalidated}. (If the observer expresses his set of probabilities in terms
of a probability density function, and a density of zero at some point
changes due to a measurement, then we can likewise say that his information
has been invalidated, since the classical conditional probability would not
cause such a change.) The basic idea of this paper is to view the projection
postulate as a ``quantum Bayes rule,'' as has also been done by Bub,$^{(2)}$
Fuchs,$^{(3)}$ and others.$^{(4)}$

In Section 2 we show how this point of view is actually an outgrowth of the
mathematics of quantum and classical mechanics formulated statistically in a
suitable (and general) way in terms of $\ast $-algebras. Mathematically
speaking, an observer's information in quantum mechanics is noncommutative,
while in classical mechanics it is commutative. Section 3 shows how the idea
of noncommutative information can clarify conceptual problems surrounding
the measuring process in quantum mechanics.

\section*{2. \textbf{PROBABILISTIC DESCRIPTION OF INFORMATION}}

In this section we present a unified statistical setting for quantum and
classical mechanics using the language of $\ast$-algebras. We consider a
physical system (quantum or classical), and an observer who can perform
measurements on the system. Our goal is essentially to describe the
observer's information regarding the system. In this vein we say that:

\begin{quote}
A \emph{measurement} on a system by an observer, is by definition the
reception of data by the observer which can change his information about the
system.
\end{quote}

Of course, we have to assume that a measurement is \emph{accurate} (i.e. the
data is correct), even though the measurement may \emph{not be precise}
(i.e. the data is not maximal), for example when the position of a classical
particle is measured, a set of possible values is obtained rather than a
single value, but the value of the position at the time of the measurement
is contained in this set.

We will view all measurements as yes/no experiments. For example, if a
particle's position is measured, and the only data the observer receives is
that the position is in the interval $[a,b]$, then we view this as a ``yes''
obtained for the yes/no experiment ``Is the particle's position in $[a,b]$%
?'' Furthermore, we assume that a measurement is ideal in the sense that a
repetition of a yes/no experiment would give the same result (assuming that
no time-evolution or other measurements takes place between the two yes/no
experiments). This is related to the accuracy of a measurement mentioned
above, since if we say that the information obtained in a measurement is
correct, it means that if we could repeat the measurement (yes/no
experiment) then we would with probability one get the same result. (We can
therefore also view a measurement as a preparation.) Note that in practice
it is not necessarily possible to repeat a yes/no experiment, since the
second measurement of position, say, might only give an interval overlapping
with the interval obtained in the first measurement, rather than giving
exactly the same interval (this would depend on the experimental setup
however). Below we will be able to give a better formal definition of an
ideal measurement. Meanwhile we note that a series of ideal measurements of
the same observable, with each measurement giving a set of values similar to
the position measurement above, should at least be consistent with each
other (assuming there's no time-evolution, measurements of other
observables, or outside influences on the system), in the sense that the
intersection of the sets obtained in the measurements should be non-empty.

\subsection*{2.1. General structure}

Here we treat the general mathematical structure of a statistical setting
for mechanics (classical and quantum). We have to give a mathematical
description of four things:

(a) The observables of the system (i.e. that which can be measured by the
observer).

(b) The state of the system (i.e. the observer's information regarding the
system). We can say that by definition the \emph{state} of the system is a
mathematical object which for each possible outcome of each measurement that
can be performed on the system, provides the observer with the probability
for obtaining that outcome when performing that measurement (these
probabilities constitute the observer's information). In the rest of this
paper, this is what the term ``state'' (of the system) will mean. The state
of the system must be constructed from measurements previously performed on
the system by the observer.

(c) The measuring process. A measurement generally changes an observer's
information about the system on which he performed the measurement, in other
words it generally changes the state of the system.

(d) The time-evolution of the system (dynamics).This describes how the
probabilities mentioned in (b) change as we move forward (or backward) in
time.

We give the following set of postulates as a mathematical description of
(a)-(d):

\bigskip(i) The observables of the system are described by a unital $\ast $%
-algebra $\frak{A}$, called the \emph{observable algebra} of the system, in
the sense that for every yes/no experiment that can be performed on the
system at a given point in time, $\frak{A}$ contains a corresponding
projection $P$, called \emph{the} projection of the yes/no experiment (at
that point in time).

\emph{Remark: }A $\ast $\emph{-algebra} is a complex vector space with a
multiplication law (that need not be commutative) and an adjoint operation $%
A\mapsto A^{\ast }$. A $\ast $-algebra is called \emph{unital} if it
contains a unit element $1$, meaning $1A=A1=A$ for all $A$ in the algebra. A 
\emph{projection} is an element $P$ having the properties $P^{2}=P=P^{\ast }$%
. See for example Refs. 5 and 6 for more on $\ast $-algebras.

\bigskip(ii) The state of the system is described by a state $\omega$ on $%
\frak{A}$ such that for every yes/no experiment, $\omega(P)$ is the
probability of getting ``yes,'' where $P$ is the projection of the yes/no
experiment at the time at which it is performed.

\emph{Remark: }A \emph{state} on $\frak{A}$ is a linear functional $\omega $:%
$\frak{A}\rightarrow\mathbb{C}$ which is positive [$\omega(A^{\ast}A)\geq0$]
and normalized [$\omega(1)=1$].

\bigskip (iii) Suppose the observer obtains a ``yes'' for a yes/no
experiment performed on the system. After the experiment the state of the
system is then given by the state $\omega ^{\prime }$\ on $\frak{A}$\
defined by 
\begin{equation}
\omega ^{\prime }(A)=\omega (PAP)/\omega (P)  \tag{1}
\end{equation}
for all $A$\ in $\frak{A}$, where $P$\ is the projection of the yes/no
experiment at the point in time at which the experiment was performed. We
can view this as the formal definition of an \emph{ideal} measurement.

\emph{Remarks:} As will be explained in Subsections 2.2, 2.3 and 2.4, Eq.
(1) is a noncommutative generalization of conditional probability. So, the
probabilities which constitute the observer's information, change according
to this (in general noncommutative) conditional probability, when the
observer performs a measurement on the system. This change, which happens in
both classical and quantum mechanics, does not refer to any influence the
measuring instruments might have on the system being measured. Such
influences are not described by the (noncommutative) conditional probability
(1), but rather by a suitable time-evolution resulting from the interaction
between system and measuring instrument, if the interaction is not
negligible.

Furthermore, in the case of commuting projections $P_{1}$ and $P_{2}$, we
have $(P_{2}-P_{1}P_{2}P_{1})^{\ast
}(P_{2}-P_{1}P_{2}P_{1})=P_{2}-P_{1}P_{2}P_{1}$ (in fact, $%
P_{2}-P_{1}P_{2}P_{1}$ is a projection), hence $\omega
(P_{2}-P_{1}P_{2}P_{1})\geq 0$, which means that $0\leq \omega
(P_{1}P_{2}P_{1})\leq \omega (P_{2})$, since $P_{1}P_{2}P_{1}=(P_{2}P_{1})^{%
\ast }(P_{2}P_{1})$. This ensures the compatibility mentioned in the
Introduction, namely that $\omega (P_{2})=0$ implies $\omega ^{\prime
}(P_{2})=0$, where $\omega ^{\prime }$ is the state given by Eq. (1) with $%
P=P_{1}$ after a ``yes'' in a yes/no experiment with projection $P_{1}$.
This always happens in the classical case, since the observable algebra is
commutative. A well-known counter example in the quantum case is a sequence
of measurements of the polarization of light.

\bigskip(iv) The time-evolution of the system is given by a one-parameter $%
\ast$-automorphism group $\tau$ of $\frak{A}$, such that if at time $0$ the
projection of a given yes/no experiment is $P$, then at time $t$ the
projection of the same yes/no experiment will be $\tau_{t}(P)$. (The choice
of when time $0$ is, is arbitrary, since $\tau$ is a group.) So we're using
the ``Heisenberg picture,'' namely the time-evolution does not act on the
state of the system.

\emph{Remark:} Mathematically this means that $\tau_{t}$ is a $\ast $%
-automorphism [a linear bijection $\frak{A}\rightarrow\frak{A}$ such that $%
\tau_{t}(AB)=\tau_{t}(A)\tau_{t}(B)$ and $\tau_{t}(A^{\ast})=\tau
_{t}(A)^{\ast}$] for each $t\in\mathbb{R}$, with $\tau_{0}$ the identity
mapping on $\frak{A}$, and $\tau_{t+s}=\tau_{t}\tau_{s}$.

\bigskip These postulates can be viewed as \emph{the general structure of
mechanics} in a statistical setting. We now briefly discuss how quantum and
classical mechanics fit into this picture.

\subsection*{2.2. Classical mechanics}

Assume that the system is classical, and that its phase space is $F$, which
we view purely as a measurable space (in the sense of measure theory; see
Ref. 7 for an exposition of the measure theoretic ideas we will use here).
The \emph{pure state} of the system is by definition the system's point in
the phase space $F$. The state of the system (in the sense of Subsection
2.1) represents the observer's information regarding the pure state. (The
pure state is the state of maximal information.)

We then take the observable algebra as $\frak{A}=B_{\infty }(F)$, which is
the unital $\ast $-algebra of all bounded complex-valued measurable
functions on $F$. The multiplication law of $\frak{A}$ is just pointwise
multiplication of functions, and the adjoint operation is simply complex
conjugation. Note that $\frak{A}$ is commutative (or abelian), in other
words $fg=gf$ for all $f,g\in \frak{A}$.

The projections of $\frak{A}$ are the characteristic functions, which we
will denote by $\chi $. For any measurable set $S\subset F$, the \emph{%
characteristic function} $\chi _{S}$ is the real-valued function on $F$
assuming the value $1$ on $S$, and zero everywhere else. A yes/no experiment
can then be viewed as having the form ``Is the pure state of the system in $%
S $?'' and the projection of this yes/no experiment is $\chi _{S}$. This
happens as follows: Any observable $f$ of the system can be represented as a
measurable function $f:F\rightarrow \mathbb{R}$. (Note that this function
can be unbounded, and is therefore not necessarily contained in $\frak{A}$.)
If $f$ is measured when the pure state is $x$, then the precise result would
be $f(x)$. A yes/no experiment concerning this observable in general has the
form ``Is the value of $f$ in $V$?'' where $V$ is some Borel set in $\mathbb{%
R}$. The projection of this yes/no experiment is $\chi _{f^{-1}(V)}$, since $%
f^{-1}(V)$ is the set in phase space corresponding to a value of $f$ in $V$.

We can describe the observer's information regarding the pure state of the
system by a probability measure $\mu $ on $F$ such that $\mu (S)$ is the
probability that the pure state is contained in the measurable set $S$. This
gives a state $\omega $ on $\frak{A}$ defined by 
\begin{equation*}
\omega (g)=\int gd\mu \text{.}
\end{equation*}
This indeed describes the system's state in the sense of postulate (ii),
since 
\begin{equation*}
\omega (\chi _{S})=\mu (S)\text{.}
\end{equation*}

For the yes/no experiment with projection $\chi_{S}$, postulate (iii) is
equivalent to the probability measure changing in the case of a ``yes'' to $%
\mu^{\prime}$ defined by 
\begin{equation}
\mu^{\prime}(U)=\mu(U\cap S)/\mu(S)  \tag{2}
\end{equation}
for any measurable set $U$ in $F$. Here $\mu^{\prime}$ is the probability
measure describing the observer's information after the yes/no experiment.
This equivalence, namely that $\mu^{\prime}$ is the (necessarily unique)
probability measure giving the state $\omega^{\prime}(g)=\int gd\mu^{\prime}$
on $\frak{A}$ satisfying Eq. (1), follows from standard measure theoretic
arguments involving Lebesgue convergence.

From Eq. (2) it is clear that in the classical case postulate (iii) is
merely a conditional probability. We therefore view postulate (iii) as a
generalization of a conditional probability to the case where the observable
algebra can be noncommutative (as is the case for quantum mechanics
discussed in Subsection 2.3).

The time-evolution of the system is given by a \emph{flow} $%
T_{t}:F\rightarrow F$ which for every $t\in \mathbb{R}$ maps measurable sets
to measurable sets, and has the group property $T_{s+t}=T_{s}T_{t}$ with $%
T_{0}$ the identity mapping on $F$ (which implies that $T_{t}$ is bijective
for every $t$). This indeed defines a one-parameter $\ast $-automorphism
group on $\frak{A}$ by the Koopman$^{(8)}$ construction 
\begin{equation*}
\tau _{t}(g)=g\circ T_{t}
\end{equation*}
which then gives the time-evolution in the sense of postulate (iv).

It should be clear that what we have described here is little more than
probability theory. Physics really only enter once we specify $F$ and $T_{t}$%
. Even so, this is how the general structure can be used as a statistical
setting for classical mechanics.

\emph{Remark:} We can in principle consider the state $\omega $ and the
projections of the various yes/no experiments of the system as being
fundamental concepts. The point in phase space is then nothing more than a
convenient theoretical construct which can be viewed as the objective state
of the system, since it is not changed by the conditional probability during
a measurement. When we move to the more general noncommutative setting of
quantum mechanics, the idea of an objective state is not possible anymore.
The noncommutative conditional probability simply does not allow it. (We
return to this in Subsection 2.3.) It therefore seems conceptually sensible
not to view the point in phase of a classical system as having fundamental
physical significance. This line of thought was also promoted by Born,$%
^{(9)} $ pp.164-170.

\subsection*{2.3. Quantum mechanics}

Assume that the system is quantum mechanical, and that its state space is $%
\frak{H}$, which is a Hilbert space (for example an $L^{2}$-space containing
wave functions). We then take the observable algebra as $\frak{A}=\frak{L(H)}
$, which is the unital $\ast $-algebra of all bounded linear operators $%
\frak{H}\rightarrow \frak{H}$. Since the observables of the system
(represented as linear operators in $\frak{H}$) can be unbounded, they are
not necessarily contained in $\frak{A}$. Rather, $\frak{A}$ contains the
spectral projections of the observables. The multiplication law of $\frak{A}$
is usual multiplication of operators, and the adjoint operation is the usual
Hilbert adjoint. Note that $\frak{A}$ is noncommutative, in other words
there are $A$ and $B$ in $\frak{A}$ such that $AB\neq BA$. (We are assuming
that $\frak{H}$ is a non-trivial state space, that is to say its dimension
is greater than one.)

We can describe the observer's information regarding the system by a density
operator (or density matrix) $\rho$ on $\frak{H}$. [By definition, $\rho$ is
an element of $\frak{L(H)}$ with the properties $\rho\geq0$ and Tr$(\rho)=1$%
, where Tr denotes the trace of an operator.] This gives a state $\omega$ on 
$\frak{A}$ defined by 
\begin{equation*}
\omega(A)=\text{Tr}(\rho A)
\end{equation*}
which describes the system's state in the sense of postulate (ii).

For the yes/no experiment with projection $P$, postulate (iii) is equivalent
to the well known projection postulate 
\begin{equation}
\rho ^{\prime }=P\rho P/\text{Tr}(\rho P)  \tag{3}
\end{equation}
of quantum mechanics, where $\rho ^{\prime }$ is the density operator
describing the observer's information after a ``yes'' was obtained for the
yes/no experiment. The equivalence follows from the fact that if Tr$(\rho
_{1}A)=$ Tr$(\rho _{2}A)$ for all $A\in \frak{L(H)}$ for two density
operators $\rho _{1}$ and $\rho _{2}$ on $\frak{H}$, then setting $A=\rho
_{1}-\rho _{2}$ gives 
\begin{equation*}
\left\| (\rho _{1}-\rho _{2})^{2}\right\| _{1}=\text{Tr}((\rho _{1}-\rho
_{2})^{2})=0
\end{equation*}
where $\left\| \cdot \right\| _{1}$ denotes the trace-class norm.$^{(6)}$
Hence $(\rho _{1}-\rho _{2})^{2}=0$ and therefore $\left\| \rho _{1}-\rho
_{2}\right\| ^{2}=\left\| (\rho _{1}-\rho _{2})^{2}\right\| =0$, where $%
\left\| \cdot \right\| $ denotes the usual operator norm. So $\rho _{1}=\rho
_{2}$, proving the equivalence, namely that $\rho ^{\prime }$ is the \emph{%
unique} density operator insuring that $\omega ^{\prime }(A)=$ Tr$(\rho
^{\prime }A)$ satisfies Eq. (1).

Our comments concerning the classical case lead us to conclude that the
projection postulate of quantum mechanics is a generalization of conditional
probability to the case of a noncommutative observable algebra (also see Bub$%
^{(2)}$). So the projection postulate is indeed a quantum Bayes rule, as
mentioned in the Introduction. Also see Ref. 10 for a short survey of the
closely related idea of noncommutative conditional expectations.

Eq. (3) of course contains the projection postulate for a state vector (and
in particular the collapse of a wave function) as a special case. [One can
in fact also work the other way around, deriving postulate (iii) from the
projection postulate for a state vector in Hilbert space, through a
heuristic argument using the GNS construction.$^{(11)}$] Since a measurement
generally changes a state vector by the projection postulate, we have to
conclude in the framework that we have set up so far, that a state vector is
observer dependent rather than being a property of the system (unlike the
pure state of a classical system, which can be viewed as having an objective
reality). In other words, even when the system's state $\omega $ is given by
a state vector $\psi \in \frak{H}$, namely $\omega (A)=\left\langle \psi
,A\psi \right\rangle $, the state is still nothing more than the observer's
information, which changes via the noncommutative conditional probability
when a measurement is made. This point of view was also taken by Fuchs and
Peres.$^{(12)}$ Another way to see it is as follows: The same density matrix
can be obtained from two different sets of state vectors (the two sets
having no scalar multiples of state vectors in common), where the vectors
from such a set is weighed by probabilities to give the density matrix.
Since a density matrix represents the observer's information, it therefore
does not make sense for the observer to say that the system has a state
vector when he does not know what the state vector is.

The time-evolution of the system is given in the sense of postulate (iv) by
the one-parameter $\ast $-automorphism group on $\frak{A}$ defined by 
\begin{equation*}
\tau _{t}(A)=e^{iHt/\hbar }Ae^{-iHt/\hbar }
\end{equation*}
where $H\,$\ is the Hamiltonian of the system.

\emph{Remark:} For an observable represented by a (possibly unbounded)
self-adjoint linear operator $A$ in $\frak{H}$, the projection of the yes/no
experiment ``Is the value of $A$ in $V$?'' can be taken as the spectral
projection $\chi _{V}(A)$ in terms of the Borel functional calculus on
self-adjoint operators; see Ref. 13, pp. 197-200 and 230, for the
construction and properties of this calculus. Loosely speaking, this
projection represents the part of $A$ whose spectrum is contained in the
Borel subset $V$ of $\mathbb{R}$. It is interesting to note that this is
very similar to the classical case in Subsection 2.2, where we used $\chi
_{f^{-1}(V)}=\chi _{V}\circ f$ instead of $\chi _{V}(A)$. We can write $\chi
_{V}(f):=\chi _{V}\circ f$ to complete the analogy, where more generally $%
g(f):=g\circ f$ defines a Borel functional calculus on the measurable
functions $f:F\rightarrow \mathbb{R}$ for Borel measurable $g:\mathbb{R}%
\rightarrow \mathbb{C}$. Furthermore, for two disjoint Borel sets $U$ and $V$
in $\mathbb{R}$, we have $\chi _{U}(A)\chi _{V}(A)=\chi _{U\cap V}(A)=0$, so
by Eq. (1) two consecutive measurements of the observable $A$ (quantum or
classical) cannot give contradicting results (namely that $A$'s value is in
both $U$ and $V$), as was mentioned just before Subsection 2.1.

The $\ast $-algebraic (or C*-algebraic) approach to quantum physics is
described at length in Haag.$^{(14)}$

\subsection*{2.4. Information}

As mentioned in Subsection 2.2, in the classical case the general structure
that we've described is actually just probability theory. One can shift the
perspective somewhat by saying that in the classical case this general
structure is a probabilistic description of information. We also saw that
quantum mechanics has exactly the same the general structure, except that it
is noncommutative. In particular, the projection postulate of quantum
mechanics is a noncommutative conditional probability. The mathematics
therefore seem to tell us that the general structure of quantum mechanics is
a mathematical framework for the probabilistic description of
``noncommutative information.'' This noncommutative nature of information in
quantum mechanics is what causes the essential difference between quantum
mechanics and classical mechanics.

We can view (i)-(iv) as the abstract axioms for a probabilistic description
of information. We can define an observer's \emph{information} as being (the
probabilities given by) the state on the observable algebra, along with a
rule describing how these probabilities change when new data is received,
namely the (noncommutative) conditional probability (iii). The information
is called \emph{noncommutative} in the case of the general noncommutative
conditional probability. If we were to add to axioms (i)-(iv) the assumption
that the observable algebra is commutative, then we get an abstract
formulation of classical probability theory with the usual conditional
probability, in which case the information can be called \emph{commutative}.
The algebras $B_{\infty }(F)$ and $\frak{L(H)}$ are nothing more than
convenient representations (of the commutative and noncommutative cases
respectively), suitable for doing physics.

\section*{3. THE MEASURING PROCESS}

Interpreting quantum mechanics as a probabilistic description of
noncommutative information, implies that an (ideal) measurement changes an
observer's information, rather than disturbing the system as is often argued
(for example in Ref. 15, p. 46). This renders many problems surrounding the
measuring process in quantum mechanics no more difficult than in classical
mechanics. The answer to both question at the beginning of this paper is
simply that the observer's information changed (i.e., the observer received
new data), exactly as for the corresponding classical questions. (In
particular this means that consciousness has no role to play in the
measuring process. The observer could be a computer connected to a measuring
instrument, or the measuring instrument itself, as long as it can obtain
information about the system.) We give a few more examples:

\bigskip\ (a) The Heisenberg cut. This refers to an imaginary dividing line
between the observer and the system being observed (see for example Ref. 16,
pp. 419-421 and 439-445, and Ref. 14, p. 295). It can be seen as the place
where information crosses from the system to the observer, but it leads to
the question of where exactly it should be; where does the observer begin?
In practice it's not really a problem: It doesn't matter where the cut is.
It is merely a philosophical question which is already present in classical
mechanics, since in the classical case information also passes from the
system to the observer and one could again ask where the observer begins.
The Heisenberg cut is therefore no more problematic in quantum mechanics
than in classical mechanics.

\bigskip (b) When does the collapse of the wave function take place and how
long does it take? This is essentially the Heisenberg cut with space
replaced by time. One can pose the question as follows: When does an
observer ``absorb'' the data (i.e., when does the measurement take place, or
when does the observer's information change), and how long does it take?
Again the quantum case is no different from the classical case, and
moreover, in practice it is no more of a problem than in the classical case.

\bigskip (c) Continuous observation (see Refs. 17 and 18 for example). The
ideal measurement discussed in Section 2 refers to a single measurement made
at some point in time. It can therefore not be applied directly to
continuous observation, i.e. when the observer's information is continually
changing. However, in classical mechanics this is not considered a
conceptual problem, since one could in principle describe such a situation
as a continual change in the probability distribution (probability measure)
describing the information, even though it might be a difficult technical
problem in practice. The same is true in quantum mechanics, with the
probability distribution replaced by a state representing noncommutative
information. (In quantum mechanics however, the idea of continuous
observation is probably an idealization, for example watching something
without blinking your eyes is not a continuous measurement, since the
photons registered by your retina are discrete.)

The ``paradox of the watched pot that never boils'' (called \emph{Zeno's
paradox} by Misra and Sudarshan$^{(19)}$) is resolved by noting that if an
observer continuously measures a certain observable, then the system can
still evolve in time to produce other values for the observable if the
measurement is not precise (as is typically the case). Say the observer
measures an observable $A$ which has a discrete spectrum, and he can only
determine its value up to some interval containing (at a point in time) a
number of eigenvalues of the observable, say $a_{1},...,a_{n}$. Then the
state vector is projected onto the subspace spanned by the eigenstates (at
that point in time) corresponding to $a_{1},...,a_{n}$, in other words, onto
the subspace which at that point in time corresponds to the interval (keep
in mind that time-evolution acts on the observable algebra, and hence on the
eigenstates of the observable). This happens according to postulate (iii);
see for example Ref. 20, pp. 260-266. To clarify our argument, we assume
here that before the continuous measurement starts, the observer has maximal
information, i.e. his information is a state vector [the general case does
not differ significantly, since it is still handled with the same projection
postulate (iii)]. Note that the state is now still a state vector, and not a
mixture of the eigenstates corresponding to $a_{1},...,a_{n}$. The interval
which is measured (and hence the eigenvalues of $A$ contained in it) can
change in the course of time (for example it can drift up and down the real
line), simply because of the lack of precision in the continuous
measurement. Therefore the value of $A$ can change within this drifting
interval, in turn allowing the drifting interval's average location to
change accordingly, which is what the observer sees. In the mathematics this
looks as follows: The continuous measurement confines the state vector via
the projection postulate to the ``drifting'' subspace corresponding to the
drifting interval. The observable's eigenstates are evolving in time, but
since this drifting subspace contains many eigenstates of the observable at
any point in time, the projection postulate does not cause the state vector
to be ``dragged along'' by one of the time-evolving eigenstates. Also, since
the interval is drifting, eigenstates are moving in and out of the subspace.
Therefore the state vector can be projected onto subspaces containing new
eigenstates (corresponding to new eigenvalues), with eigenstates brought
closer to the state vector by time-evolution having higher probability.
(This argument becomes somewhat clearer in the Schr\"{o}dinger picture,
where the eigenstates are fixed, but the subspace is still drifting.)

If the continuous measurement is precise enough, then quantum mechanics
indeed predict that ``a watched pot never boils'' if the observable's
eigenvalues are discrete (precise measurement of a continuous observable is
impossible in practice). This happens because a quantum measurement can
invalidate previous information (i.e. the state vector can change by
projection) which then ``cancels out'' the changes due to time-evolution
acting on the observable algebra (and thus on the observable's eigenvectors
onto which projection of the state vector occurs). In effect the state
vector is dragged along by the time-evolving eigenstate corresponding to the
measured value. In classical mechanics on the other hand, previous
information is not invalidated by measurement, hence the pure state of a
system will not change because of continuous observation, and therefore the
values of observables can change as time-evolution acts on the observable
algebra. Note that this is true even if the classical observable being
observed is discrete (for example ``number of particles in the left half of
the container''). So no matter how closely we watch a classical pot, it can
still boil.

\bigskip (d) The EPR ``paradox.'' Einstein, Podolsky and Rosen$^{(21)}$
described a now famous experiment in which two particles are created
together (or interact) and then move away from each other (which ends any
interaction between them) before a measurement is performed on one of the
particles. This measurement then gives corresponding data about the other
particle as well. [This is the result of an entanglement of the two
particles' states (for example due to a conservation law), which can occur
since the state space is the tensor product of the two particles' state
spaces.] EPR\ argued that this means that the second particle simultaneously
has values for two noncommuting observables like position and momentum,
since only the first particle is measured (either its position or its
momentum is measured, but not both), and hence quantum mechanics must be
incomplete, since it says that a particle does not simultaneously have
values for position and momentum. They based this on the idea that a
measurement on the first particle does not disturb the second. However, we
have viewed a measurement as the reception of data by the observer; it has
nothing to do with the observer ``directly'' observing (and disturbing) the
system. Measuring the first particle gives the observer data regarding the
second particle as well (and hence \emph{is} a measurement of the second
particle), which is mathematically described by the second particle's state
vector (representing the observer's \emph{noncommutative} information about
this particle) now being in an eigenspace of the observable which was
measured. This is no different from the analogous situation in classical
mechanics where for example conservation of momentum can give the second
particle's momentum when the first particle's momentum is measured, except
that in this case information is commutative.

We can even have two observers A and B measuring the same observable of the
two particles respectively. A's measurement is then also a measurement of
the value B will get (A receives \ data about what B's result will be) and
so there's nothing strange in them getting correlated results (say opposite
values for momentum). No signal need travel faster than the speed of light
to B's particle to ``tell'' it to have the opposite value to A's result, in
the same way that no such signal is needed in the classical case. From A's
point of view, B is part of the system along with the two particles, and so
this experiment is really no different from the original one observer EPR\
experiment above. The particles \emph{along with} B are in a superposition
of states from A's point of view until A measures his particle, which
reduces (by projection) the state vector of the combined system of particles
and B, with B then in the eigenspace ``B gets the opposite value.''

\bigskip (e) System and observer as a combined system (see Ref. 22, pp.
175-183, for a short and clear discussion). Here the time-evolution of the
combined system is supposed to account for the projection postulate of
quantum mechanics. This is not possible in a natural way, since
time-evolution is the result of a one-parameter $\ast $-automorphism group.
In classical mechanics the combined system evolves according to classical
dynamics (the observer being thought of as a classical system in this case),
and this then similarly would have to account for the change in the
observer's information via a conditional probability due to a measurement he
performs on the system. Again this is not possible in a natural way, since
here too we have the same projection postulate, namely the conditional
probability (1), acting on the state (of the system without observer), while
the time-evolution acts as a one-parameter $\ast $-automorphism group on the
observable algebra. The solution is that the state of the combined system
has to contain from the start the fact that the observer will perform a
measurement on the system at a given point in time and will subsequently
experience a change of information (this change is a physical process in the
observer, described by the combined system's time-evolution, for example
some neural activity in a human observer's brain), otherwise such a
measurement and the change of information would not take place. This is
clear, since time-evolution does not act on the state, but on the observable
algebra, hence the state of the combined system is the state ``for all
time'' and does not change when the observer performs a measurement. Exactly
the same is true for quantum mechanics (where the observer is then also
viewed as a quantum system). The (noncommutative) conditional probability,
that is to say the projection postulate, is only relevant when the observer
is not considered to be part of the system, in which case the conditional
probability says what the change in the observer's information will be, it
does not describe the physical process taking place in the observer to
accommodate (or store) the new information.

\bigskip The point we attempt to make with these examples is that, even
though there might be certain problems surrounding the measuring process,
quantum mechanics does not introduce any conceptual problems not already
present in classical mechanics, as long as we assume that information is
noncommutative in quantum mechanics.

In connection with the two-slit experiment we can mention the following:
Assume that the probability density function for the position of detection
of a particle on the screen in the two-slit experiment is given by an
interference pattern when no measurement is performed at the two open slits.
This density function represents the observer's information about where on
the screen the particle will be detected. In the light of our discussion
thus far, it should then not be too surprising that this density function
(i.e. the observer's information) can be invalidated via the noncommutative
conditional probability (1), if the observer does measure through which slit
the particle goes (i.e. if the observer receives new data), giving a
completely different probability density function at the screen, for example
points that had zero probability density now having positive probability
density, as mentioned in the Introduction. This is unlike the classical case
where a measurement at the slits gives the observer more information, rather
than invalidating previous information. (Also see Ref. 2.)

We can also consider the case of more than one observer touched upon in (d).
Say three observers A, B and C are observing the same system, but B and C
are not aware of each other or of A. B and C measure two noncommuting
observables $P$ and $Q$ respectively, in the order $P$, $Q$, $P$, and A in
turn measures B and C's results in this order ( he ``sees'' each of their
results at the time they obtain them). We ignore the time-evolution of the
system. Say the results are $p_{1}$, $q$, $p_{2}$ (in this order), then
clearly $p_{1}$ and $p_{2}$ need not be the same since $P$ and $Q$ do not
commute. So from B's point of view it seems that something disturbed the
system between his two measurements of $P$. However, in our interpretation
it is actually B's information that has been invalidated by A and C's
measurement of $Q$. This is not too strange, since B and C are merely A's
measuring instruments. One could ask what would happen if A wasn't there.
Would B then get $p_{1}=p_{2}$ with probability one? In the absence of A,
does it even make sense to talk of the time order $P$, $Q$, $P$ if B and C
are not aware of each other? In our interpretation time ordering should
probably be viewed as in some way defined by events where data is received
by an observer, and in this case it seems possible that B would get $%
p_{1}=p_{2}$ with probability one in the absence of A and no other way to
define the time ordering. (Note that in the two-slit experiment for example,
there is a time ordering in the sense that a measurement on a particle at
the slits is performed before a measurement on the same particle at the
screen, even if the measurements are performed by two different observers
not aware of each other, so the interference pattern at the screen can still
be destroyed in this setup.) The idea of defining time ordering in terms of
a series of events (an event in our case being the reception of data by an
observer, or alternatively, a change in an observer's information) was
introduced by Finkelstein.$^{(23)}$

\section*{4. DISCUSSION}

We have now seen that the general structure of quantum mechanics is actually
a mathematical framework for handling noncommutative information, rather
than being a physical theory in itself.

If we assume that information in our physical world is described by quantum
mechanics, this leads us to conclude that information is actually a
noncommutative phenomenon. Perhaps this means that since information
``lives'' in spacetime (and possibly in some way defines spacetime structure
as was alluded to at the end of Section 3), spacetime itself is
noncommutative, as has been suggested in attempts to construct quantum
spacetime and quantum gravity; see for example Ref. 24. On the other
extreme, the term ``noncommutative information'' may be a ``purely
grammatical trick'' of the sort Marsden,$^{(25)}$ p. 188, mused might ``be
the ultimate solution of the quantum measurement problem''; this possibility
seems somewhat less interesting however.

It also explains the linearity of quantum mechanics. The general structure
of classical mechanics in Section 2 is linear, since it is nothing more than
probability theory, even though it can be applied to physical systems where
nonlinear aspects might be involved. It is the statistical point of view
that makes everything linear (this boils down to the use of averages, which
are integrals and hence linear). The same goes for quantum mechanics. Its
linear structure should not be viewed as an approximation to an underlying
nonlinear world, but simply as a result of the fact that it is a
mathematical framework for probability theory (i.e. statistics, averages),
where the information involved happens to be noncommutative. The appearance
of a Hilbert space as the state space is simply a mathematical way of
representing the noncommutative $\ast $-algebraic general structure in
Section 2. So the linearity of (and hence superpositions in) the state space
is just a convenient way to express the fact that a measurement can
invalidate the information the observer had before the measurement, or more
precisely, to express noncommutative conditional probabilities. Also see
Ref. 14, p. 309, and Ref. 26, p. 175, for similar remarks concerning the
linearity of quantum mechanics.

A review of quantum mechanics viewed as a generalization of classical
probability theory can be found in Ref. 27.

We cannot claim that this noncommutative information interpretation solves
all the conceptual problems of quantum physics, but for the general
situation of a quantum mechanical system being observed by an observer, it
does seem to clarify many issues without causing any new problems (except if
you consider the idea of noncommutative information itself to be a problem).

\section*{ACKNOWLEDGMENTS}

I thank the National Research Foundation, as well as the Mellon Foundation
Mentoring Programme at the University of Pretoria, for financial support.

\section*{REFERENCES}

\begin{enumerate}
\item  W. Heisenberg, \textit{The physical principles of the quantum theory}
(University of Chicago Press, Chicago, 1930).

\item  J. Bub, ``Von Neumann's projection postulate as a probability
conditionalization rule in quantum mechanics,'' \textit{J. Phil. Logic }%
\textbf{6}, 381-390 (1977).

\item  C. A. Fuchs, ``Quantum foundations in the light of quantum
information,'' in \textit{Proceedings of the NATO Advanced Research Workshop
on Decoherence and its Implications in Quantum Computation and Information
Transfer}, edited by A. Gonis (Plenum, New York, 2001), \texttt{%
quant-ph/0106166}, \texttt{quant-ph/0205039}.

\item  C. M. Caves, C. A. Fuchs, and R. Schack, ``Quantum probabilities as
Bayesian probabilities,'' \textit{Phys. Rev. A }\textbf{65}, 022305 (2002), 
\texttt{quant-ph/0106133}.

\item  O. Bratteli and D. W. Robinson, \textit{Operator Algebras and Quantum
Statistical Mechanics 1}, 2nd ed. (Springer-Verlag, New York, 1987).

\item  G. J. Murphy, \textit{C*-algebras and operator theory} (Academic
Press, San Diego, 1990).

\item  W. Rudin, \textit{Real and complex analysis}, 3rd ed. (McGraw-Hill,
1987).

\item  B. O. Koopman, ``Hamiltonian systems and transformations in Hilbert
space,'' \textit{Proc. Natl. Acad. Sci. U.S.A.} \textbf{17}, 315-318 (1931).

\item  M. Born, \textit{Physics in my generation} (Pergamon, London, 1956).

\item  D. Petz, ``Conditional expectation in quantum probability,'' in 
\textit{Quantum Probability and Applications III}, edited by L. Accardi and
W. von Waldenfels (Springer-Verlag, Berlin, 1988), pp. 251-260.

\item  R. Duvenhage, ``Recurrence in quantum mechanics,'' \textit{Int. J.
Theor. Phys.\ }\textbf{41}, 45-61 (2002),\texttt{\ quant-ph/0202023}.

\item  C. A. Fuchs and A. Peres, ``Quantum theory needs no
`interpretation','' \textit{Phys. Today} \textbf{53}(3), 70-71 (2000).

\item  S. Str\u{a}til\u{a} and L. Zsid\'{o}, \textit{Lectures on von Neumann
algebras} (Editura Academiei, Bucure\c{s}ti and Abacus Press, Tunbridge
Wells, 1979).

\item  R. Haag, \textit{Local Quantum Physics: Fields, Particles, Algebras},
2nd ed. (Springer-Verlag, Berlin, 1996).

\item  J. Schwinger, \textit{Quantum Mechanics: Symbolism of Atomic
Measurements} (Springer-Verlag, Berlin, 2001).

\item  J. von Neumann, \textit{Mathematische Grundlagen der Quantenmechanik}
(Springer, Berlin, 1932; English transl. \textit{Mathematical Foundations of
Quantum Mechanics }by R. T. Beyer, Princeton University Press, Princeton,
New Jersey, 1955).

\item  T. Sudbery, ``Continuous state reduction,'' in \textit{Quantum
Concepts in Space and Time}, edited by R. Penrose and C. J. Isham (Oxford
University Press, Oxford, 1986), pp. 65-83.

\item  A. S. Holevo, ``Limit theorems for repeated measurements and
continuous measurement processes,'' in \textit{Quantum Probability and
Applications IV}, edited by L. Accardi and W. von Waldenfels
(Springer-Verlag, Berlin, 1989), pp. 229-255.

\item  B. Misra and E. C. G. Sudarshan, ``The Zeno's paradox in quantum
theory,'' \textit{J. Math. Phys. }\textbf{18}, 756-763 (1977).

\item  C. Cohen-Tannoudji, B. Diu, and F. Lalo\"{e}, \textit{Quantum
mechanics}, Volume I (Hermann, Paris and John Wiley \& Sons, New York, 1977).

\item  A. Einstein, B. Podolsky, and N. Rosen, ``Can quantum-mechanical
description of physical reality be considered complete?,'' \textit{Phys. Rev.%
} \textbf{47}, 777-780 (1935).

\item  C. J. Isham, \textit{Lectures on Quantum Theory: Mathematical and
Structural Foundations} (Imperial College Press, London, 1995).

\item  D. Finkelstein, ``Space-time code,'' \textit{Phys. Rev.} \textbf{184}%
, 1261-1271 (1969).

\item  S. Doplicher, K. Fredenhagen, and J. E. Roberts, ``The quantum
structure of spacetime at the Planck scale and quantum fields,'' \textit{%
Commun. Math. Phys.} \textbf{172}, 187-220 (1995).

\item  J. Marsden, \textit{Application of global analysis in mathematical
physics}, Carleton Mathematical Lecture Notes No. 3 (1973).

\item  D. R. Finkelstein, \textit{Quantum Relativity: A Synthesis of the
Ideas of Einstein and Heisenberg} (Springer-Verlag, Berlin, 1996).

\item  R. F. Streater, ``Classical and quantum probability,'' \textit{J.
Math. Phys.} \textbf{41}, 3556-3603 (2000).
\end{enumerate}

\end{document}